\documentclass[%
 reprint,
 amsmath,amssymb,
 aps,
]{revtex4-1}
\usepackage{afterpage}
\usepackage{color}
\usepackage{graphicx}
\usepackage{dcolumn}
\usepackage{bm}

\newcommand{\comment}[1]{}

\begin{document}

\preprint{APS/123-QED}

\title{Hole Properties On and Off Magnetization Plateaus in 2-d Antiferromagnets}

\author{Imam Makhfudz and Pierre Pujol}
\affiliation{%
Laboratoire de Physique Th\'{e}orique--IRSAMC, CNRS and Universit\'{e} de Toulouse, UPS, F-31062 Toulouse, France 
}%
\date{\today}

\begin{abstract}

The phenomenon of magnetization plateaus in antiferromagnets under magnetic field has always been an important topic in magnetism.
We propose to probe the elusive physics of plateaus in 2-d by considering hole-doped antiferromagnet
and studying the signatures of magnetization plateaus in terms of the properties of holes, coupled to an effective gauge field generated by the spin sector. 
The latter mediates interaction between the holes, found to be algebraically decaying long-ranged
with both Coulombic and dipolar forms outside plateau and short-ranged (local) inside plateau. The resulting hole spectral weight is significantly broadened off-plateau, 
while it remains sharply-peaked on-plateau. We also extend the result obtained for 1-d system where finite hole doping gives rise to a shift in the magnetization value of the plateaus.




\end{abstract}

\pacs{Valid PACS appear here}
\maketitle


\textit{Introduction.\textemdash} Antiferromagnets under magnetic field have been known to display magnetization plateaus. The theory of
this magnetization plateaus has been an important problem in magnetism and is mainly aimed at providing explanation for such magnetization 
plateaus. Even more intriguing question is what happens if we hole-dope the antiferromagnet by removing some spins.
Hole-doped antiferromagnets have drawn much attention since the discovery of high $T_c$ Cuprate superconductivity obtained upon
hole doping the parent compound antiferromagnets \cite{DopingMott}. Most studies in this context considered Hubbard types of model at zero field
analyzed using slave-particle formalism with emergent gauge field. The topic constitutes a fundamental problem of importance to all areas of physics: matter-gauge field interaction.

In the area of magnetism itself, antiferromagnets under magnetic field are widely studied as the field helps select well defined ground state, thus allowing for the use of semiclassical approach,
and gives rise to plateaus. Magnetization plateaus are enhanced by geometric frustration and are also related to exotic states of matter, such as 
spin liquid states \cite{Natcomm}. However, most studies so far considered undoped antiferromagnets with hole-doped case not much explored 
in realistic models \cite{plateauexactsolvablemodel}.
The magnetization plateaus should have immediate consequences on the properties of holes and this is 
what we investigate in this work.

The theory of magnetization plateaus, without hole doping, can be relatively well understood with a spin path-integral approach \cite{TTH}. In one dimension it 
gives
rise to plateaus quantization condition derived based on Lieb-Schultz-Mattis theorem \cite{LMS}\cite{AffLieb} as shown first in \cite{TotsukaPLA}\cite{OshikawaAffleck}. 
The presence of holes in 1-d can also be treated with bosonization \cite{boson1}\cite{boson2}\cite{boson3}\cite{boson4} and spin path integral \cite{Shankar}\cite{CL-SC-PP}.
However, generalization of the theory to two and higher dimensions remains a challenge.
In this work, we will show that one can gain important insights into the physics of magnetization plateau 
in higher dimensions by working out the fermion-gauge field theory of hole-doped antiferromagnet.
We will demonstrate that the on and off-plateau states of antiferromagnet give rise to distinct 
types of interaction between holes and the resulting spectral function.

\textit{Field Theory.\textemdash} We employ semiclassical path integral theory of spin system \cite{TTH} and start with Euclidean space-time effective action of 2-d antiferromagnet in the presence of holes

\begin{equation}\label{spinsectoraction}
 S_{\phi}=\int d^2 x \int d\tau \frac{K_{\tau}}{2}(\partial_{\tau}\phi)^2 + \frac{K_{r}}{2}(\nabla\phi)^2 + i\left(\frac{S-m}{a^2}\right)\partial_{\tau}\phi
\end{equation}
\begin{equation}\label{holeaction}
 S_{\overline{\psi},\psi}=\int_{\mathbf{x},\tau}\overline{\psi} (\partial_{\tau}-ie A_{\tau})\psi+\int_{k's} \overline{\psi}_k \epsilon_{k's}(\phi_{k'}) \psi_{k''}\delta(\sum_{k's})
\end{equation}
describing low-energy long-distance fluctuations around classical ground state specified by $\mathbf{S}=S(\sin\theta_0\cos\phi_0,\sin\theta_0\sin\phi_0,\cos\theta_0)$ 
with spin $S$ and $z$ magnetization $m=S\cos\theta_0$,
where $\phi$ is the phase angle fluctuation field around $\phi_0$ \cite{TTH}.
The $K_{\tau},K_{r}$ are stiffness coefficients which can be determined from microscopic spin model \cite{stiffness}, giving boson velocity $v_b=\sqrt{K_r/K_{\tau}}$.
The $\overline{\psi},\psi$ represent the creation and annihilation operator fields of the (spinless) fermionic holes. The $\epsilon_{k's}(\phi_{k'})$ is fermion energy dispersion that 
couples the holes to the spin sector represented by $\phi$ field \cite{notation} via the gauge field given as $eA_{\mu}=e_g\partial_{\mu}\phi,\mu=\tau,x,y$ with $e\equiv e_g$ the effective gauge charge of the $U(1)$
gauge theory \cite{gaugecoupling}.  
Our theory will be very generic, but it is aimed to be a paradigm for spin systems well described by Heisenberg model with strong anisotropy and is under magnetic field,
\begin{equation}\label{latticespinmodel}
 H=J\sum_{ij} \mathbf{S}_i\cdot\mathbf{S}_j+D\sum_i (S^z_i)^2 - h \sum_i S^z_i
\end{equation}
with classical ground state characterized by $\cos\theta_0=\frac{h}{2S(4J+D)}$ \cite{TTH}, such as those systems with $S=3/2$ where $1/3$ plateau is expected to occur at large enough $D$ \cite{S=3per2}.

We will consider a model for holes which in the realistic case of finite doping has linear energy dispersion around Fermi surface.
The hole doping itself will give feedback effect to the spin sector. In such linear fermion dispersion, a sea of occupied negative energy states 
arises due to linearization and must be removed by applying projection operator \cite{Shankar}\cite{CL-SC-PP}; $P_j=1-\psi^{\dag}_j\psi_j/(2S)$ at each site $j$ on the microscopic lattice model.
The doping in turn modifies the plateau quantization condition via normal ordering of the fermion bilinear operator; $\psi^{\dag}_j\psi_j=\delta+ :\psi^{\dag}_j\psi_j:$, 
where $\delta=\langle \psi^{\dag}_j\psi_j\rangle$ is the doping level. We find that with hole doping $\delta$, plateau occurs at

\begin{equation}
 \left(1-\frac{\delta}{2S}\right)\left(S\pm m\right)\in \mathbb{Z}
\end{equation}
indicating a shift in magnetization plateu, proportional to doping level $\delta$, compared to the zero doping case, confirming the result in 1-d \cite{CL-SC-PP}.

As was shown in \cite{TTH}, the presence of the Berry phase term plays a crucial role in the large scale physics of the spin sector.
If the factor in front of it is an arbitrary real number, field configuration with vortices are forbidden by quantum interference and the Goldstone field 
$\phi$ is protected and the system does shown long range order and gapless behavior with no plateau. On the contrary, when the Berry phase factor
is an integer, vortex configurations are allowed and, for some values of the spin field stiffness, the system may disorder and acquire a gap. This
is the plateau situation which can phenomenologically represented by an effective mass term in the Goldstone field, writable as $m^2_{\phi}\phi^2/2$, into the effective action
Eq. (\ref{spinsectoraction}). 

We describe holes in antiferromagnet as follows.
For concreteness, we consider a simple model with holes hopping on square lattice with nearest-neighbor tight-binding dispersion $\epsilon^0_k=-2t(\cos k_x + \cos k_y)-\mu$. 
This gives Fermi surface with shape which depends on the chemical potential (and thus filling factor);
at chemical potential $\mu=-4t$ we get a Fermi point (corresponding to zero or thermodynamically small number of hole doping), 
at $-4t<\mu<0$ we get roughly circular Fermi surface that can be described by $k^2_{Fx}+k^2_{Fy}=k^2_F=4+\mu/t$ 
and at half filling $\mu=0$, we get a square-shaped Fermi surface described by $k_{Fy}=\pm k_{Fx}\pm \pi$.

\begin{figure}
 \centering
a)\includegraphics[scale=0.140]{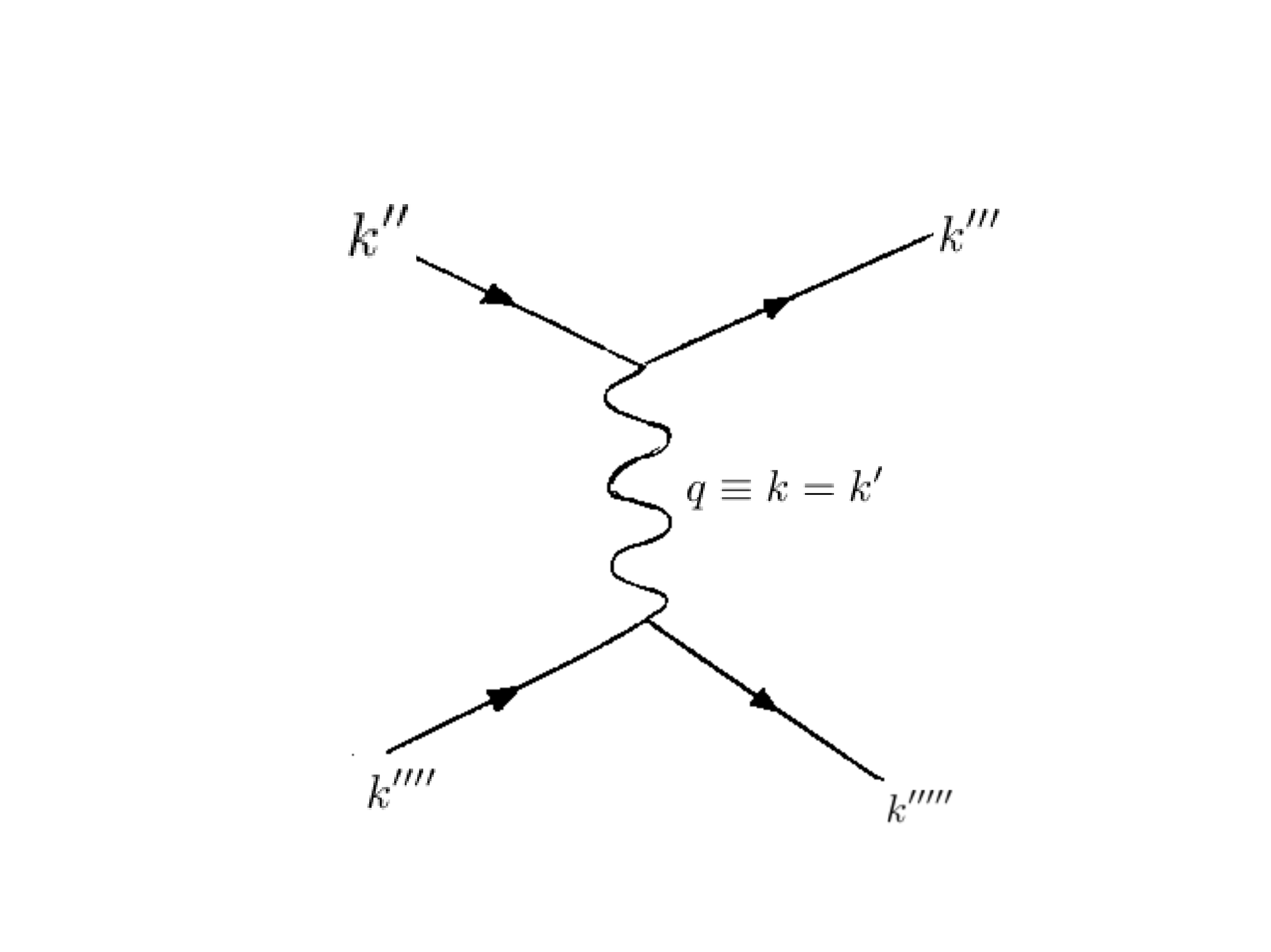}
b)\includegraphics[scale=0.140]{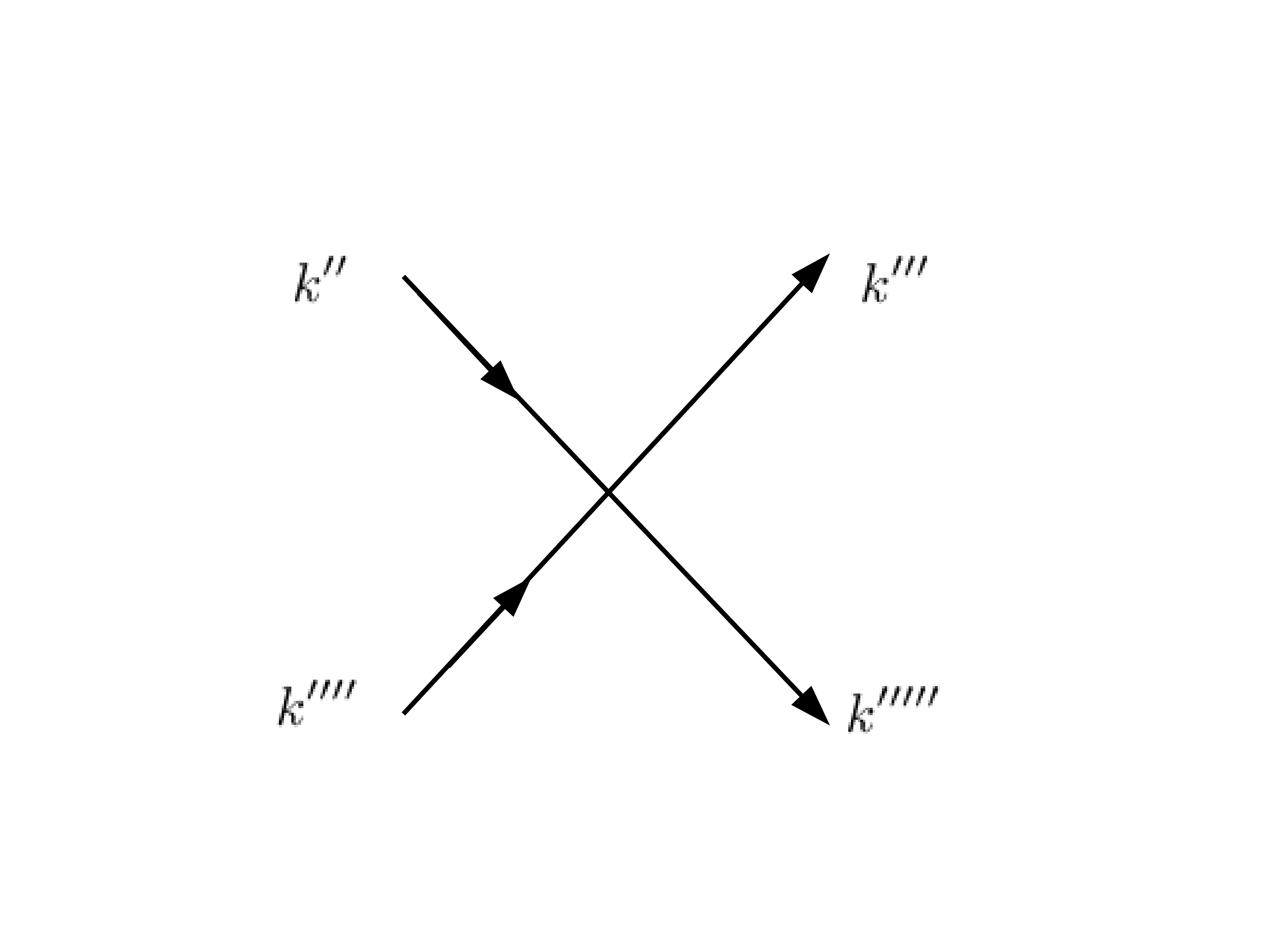}
\caption{a) Feynman diagram of 2-fermion scattering process mediated by gauge field b) The local interaction vertex counterpart.}
\label{fig:2-fermion}
\end{figure}

The fermionic holes will be coupled to gauge field generated by spin sector. 
An effective action for hole with such coupling can be derived   
by considering tight-binding hopping Hamiltonian \cite{CL-SC-PP} with hopping integral which involves the overlap of the spin coherent states at 
the neighboring sites between which the hole hops \cite{Shankar}, giving the spatial part of gauge field $A_x,A_y$, plus applying projection operator that represents the process of doping holes \cite{CL-SC-PP},
giving the temporal part of gauge field $A_{\tau}$. The result is equivalent to a minimal coupling $-i\partial_{\mu}\rightarrow -i\partial_{\mu}-eA_{\mu}$ between the spin sector's gauge field and the hole.
Considering nearest-neighbor tight binding Hamiltonian on square lattice and applying this minimal coupling to the free hole dispersion $\epsilon^0_k$ gives
$\epsilon_{k's}(\phi_{k'})=-2t(\cos (k_x-ie_{g}k'_x\phi_{k'}) + \cos (k_y-ie_{g}k'_y\phi_{k'}))$.
Performing Taylor expansion to the two cosine terms around the minimum of the band and doing the Euclidean space-time functional integral, we obtain
\begin{equation}
 Z=\int D\overline{\psi}D\psi e^{-\int_k\overline{\psi}_k\epsilon^0_k\psi_k-\delta S_{\overline{\psi},\psi}}
\end{equation}
where
\begin{equation}\label{holeactionrenormalizationKspace0}
\delta S^{\mathrm{quadratic}}_{\overline{\psi},\psi}=\int_{k's}\overline{\psi}_{k'''}\psi_{k''} F(k's)G_{\phi}(k) \overline{\psi}_{k'''''}\psi_{k''''}\delta(\sum_{k's})
\end{equation}
where the function $F(k's)$ and the propagator $G_{\phi}(k)$ of Goldstone field are given by 
\begin{equation}
 F(k's)=e^2_g[\frac{1}{4}k^2_0+t^2(\mathbf{k}\cdot\mathbf{k}''')(\mathbf{k}\cdot\mathbf{k}''''')]
\end{equation}
\begin{equation}
 G^{-1}_{\phi}(k)=[\frac{K_{\tau}}{2}k^2_0 + \frac{K_{r}}{2}\mathbf{k}^2+\frac{1}{2}m^2_{\phi}]
\end{equation}
in Euclidean space-time \cite{SuppMat}.
We see that the main effects of the spin sector manifest in the form of 4-fermion interaction term (scattering between two fermions)
with kernel which is massless for long-range interaction between vortex loops but gapped for short-range interaction between vortex loops.
We note that outside the plateau where $m_{\phi}\rightarrow 0$, as $|\mathbf{k}|\rightarrow 0$ the kernel goes as $k_{\alpha}G_{\phi}(k)k_{\beta}\sim 1/K_{\tau,r}$,
while in the plateau where $m_{\phi}\rightarrow \infty$, the kernel goes as $k_{\alpha}G_{\phi}(k)k_{\beta}\rightarrow 0$. 
This implies that within the plateau, we have true short-range interaction between fermionic holes whereas outside the plateau, 
we have nonlocal algebraically decaying interaction between fermionic holes \cite{Coulomb}. This 2-fermion scattering action 
is best illustrated by the Feynman diagram in Fig.~\ref{fig:2-fermion}a) \cite{PS-QFT}.

An important result of this work is the final form of this 4-fermion interaction term in the out-of-plateau and in-plateau cases, which in Euclidean space-time can be written as

\begin{equation}\label{kernelgeneral}
 \delta S^{\mathrm{quadratic}}_{\overline{\psi},\psi}=\int_{k's}\overline{\psi}_{k'''''}\psi_{k''''}V(k's)\overline{\psi}_{k'''}\psi_{k''}\delta(\sum_{k's})
\end{equation}
where
\begin{equation}\label{kernelconf}
 V^{\mathrm{out-of-plat}}(k's)=\frac{2}{K_s}e^2_g\frac{[t^2(\mathbf{q}\cdot\mathbf{k}''')(\mathbf{q}\cdot\mathbf{k}''''')+\frac{1}{4}q^2_0]}{q^2_0 + \mathbf{q}^2+\tilde{m}^2_{s}}
\end{equation}
\[
  V^{\mathrm{in-plat}}(k's)=\frac{2}{K_s}Ae^2_g[-t^2(\mathbf{k}'''\cdot\mathbf{k}''-\mathbf{k}''^2)(\mathbf{k}'''''\cdot\mathbf{k}''''-\mathbf{k}''''^2)
\]
\begin{equation}\label{kerneldeconf}
  -\frac{1}{4}(k'''''_0-k''''_0)(k'''_0-k''_0)]
\end{equation}
for the out-of-plateau ($m_{\phi}\rightarrow 0$) and in-plateau ($m_{\phi}\rightarrow \infty$) cases represented in Fig.~\ref{fig:2-fermion}a) and b), respectively.
We have rescaled $K_{\tau}=K_r=K_s$ (equivalent to setting the boson velocity to unity $v_b=1$) and the constants are $\tilde{m}^2_s=m^2_{\phi}/K_s$
and $A=1/\tilde{m}^2_{s}$ \cite{SuppMat}. Interestingly, $V^{\mathrm{out-of-plat}}(k's)$ contains algebraically decaying 
interaction with dipolar form in real space in addition to the more conventional density-density interaction term, 
\begin{equation}\label{dipolarkernel}
\delta S_{dip}=\frac{2}{K_s}t^2e^2_g\int_{x_{\mu},x'_{\mu}}\frac{3(\mathbf{d}_1.\Delta\mathbf{r})(\mathbf{d}_2.\Delta\mathbf{r})-(\mathbf{d}_1\cdot\mathbf{d}_2)|\Delta x|^2}{4\pi|\Delta x|^5}
\end{equation}
\begin{equation}\label{density-density}
\delta S_{dens}=-\frac{1}{2K_s}e^2_g\int_{x_{\mu},x'_{\mu'}}\rho(x)\left(\frac{3(\Delta\tau)^2-|\Delta x|^2}{4\pi|\Delta x|^5}\right)\rho(x')
\end{equation}
with dipole moments $\mathbf{d}_1=[\nabla\overline{\psi}(x)]\psi(x),\mathbf{d}_2=[\nabla\overline{\psi}(x')]\psi(x')$,$\Delta\tau=\tau'-\tau$,$\Delta\mathbf{r}=|\mathbf{x}'-\mathbf{x}|$, and $|\Delta x|=|x'_{\mu}-x_{\mu}|$, 
where $x_{\mu}=(\tau,\mathbf{x})$, $\rho(x)=\overline{\psi}(x)\psi(x)$. In each of Eqs. (\ref{kernelconf}) and (\ref{kerneldeconf}), the spatial momentum part represents the dipole interaction 
whereas the frequency ($k_0$'s) part represents the density-density interaction. Surprising as it is, dipolar interaction intuitively originates from spatial nonuniformity of the hole density distribution, 
which gives rise to nonzero effective dipole moment, corresponding to nonzero Fourier wavevectors $\mathbf{k}'s\neq 0$. 
Such dipolar term will vanish for spatially uniform distribution of holes, where only $\mathbf{k}'s=0$ remains.
The presence of both space and time distances in Eqs. (\ref{dipolarkernel}) and (\ref{density-density}), 
corresponding to the presence of both momentum and frequency dependences in the kernel Eq. (\ref{holeactionrenormalizationKspace0}), reflects the fact that the long-range interaction is not instantaneous as it is mediated by Goldtsone bosons 
with low speed $v_b\ll c$ in reality. $S_{dens}$ has asymptotic spatial dependence $V(r)\sim1/r^3$ at large distances and is repulsive \cite{SuppMat}. 
This unexpected result arises from the peculiarity of the gauge field with its origin from the spin sector's physics and its coupling to holes. 

Next, we consider finite but low doping levels at $-4t< \mu< 0$ , where we have roughly a circular Fermi surface. 
In this case, we obtain linearized dispersion $\epsilon_k=2t[(k_x-k_{xF})\sin k_{xF}+(k_y-k_{yF})\sin k_{yF}]$ 
where $k^2_{xF}+k^2_{yF}=k^2_F=4+\mu/t$ derived using Taylor series expansion of nearest-neighbor tight-binding energy dispersion $\epsilon_k=-2t(\cos k_x+\cos k_y)-\mu$ around Fermi surface satisfying
$-2t(\cos k_{xF}+\cos k_{yF})-\mu=0$  \cite{circFSdisp}.
We obtain 4-fermion interaction and kernels similar to Eqs. (\ref{holeactionrenormalizationKspace0}),(\ref{kernelconf}),(\ref{kerneldeconf})
but with the integrals over fermion momenta constrained to be near Fermi surface only \cite{SuppMat}.
We observe that the distinction of the physics of the spin sector on and off plateau manifests in the form of distinct fermion-fermion interaction between holes in hole-doped antiferromagnet. 

\textit{In vs. Out of Plateau Physics from Hole Properties.\textemdash}
We will consider the signature of these different types of interactions arising from the in-plateau and out-of-plateau states of spin sector in terms of fermion Green's function renormalization 
and spectral function of holes, which is a quantity typically measured by photoemission experiment when one is interested in the charge degree of freedom. 
The spectral function, a generalization of the density of states, is defined as
where $G(\mathbf{k},\omega)$ is renormalized Green's function which embodies the effects of interaction of fermions with each other and with other degrees of freedom. 
We will consider approximation where we geometric sum a particular family of diagrams
involving series of one-loop fermion self-energy diagrams and obtain the familiar result, 
where in this case $\Sigma(\mathbf{k},\omega)$ is the one-loop self-energy correction to free fermion Green's function $G^{-1}_0(\mathbf{k},\omega)=\omega-\epsilon_{\mathbf{k}}+i\eta \mathrm{sgn}(|\mathbf{k}|-k_F)$
with $\eta$ an infinitesimally small positive number to be taken to zero at the end of calculation \cite{AGD}.
The distinction in the profile of hole spectral function is what we expect to be a prospective experimental signature that distinguishes the physics of antiferromagnet between within and outside of plateau.

\begin{figure}
  \centering
\includegraphics[scale=0.15]{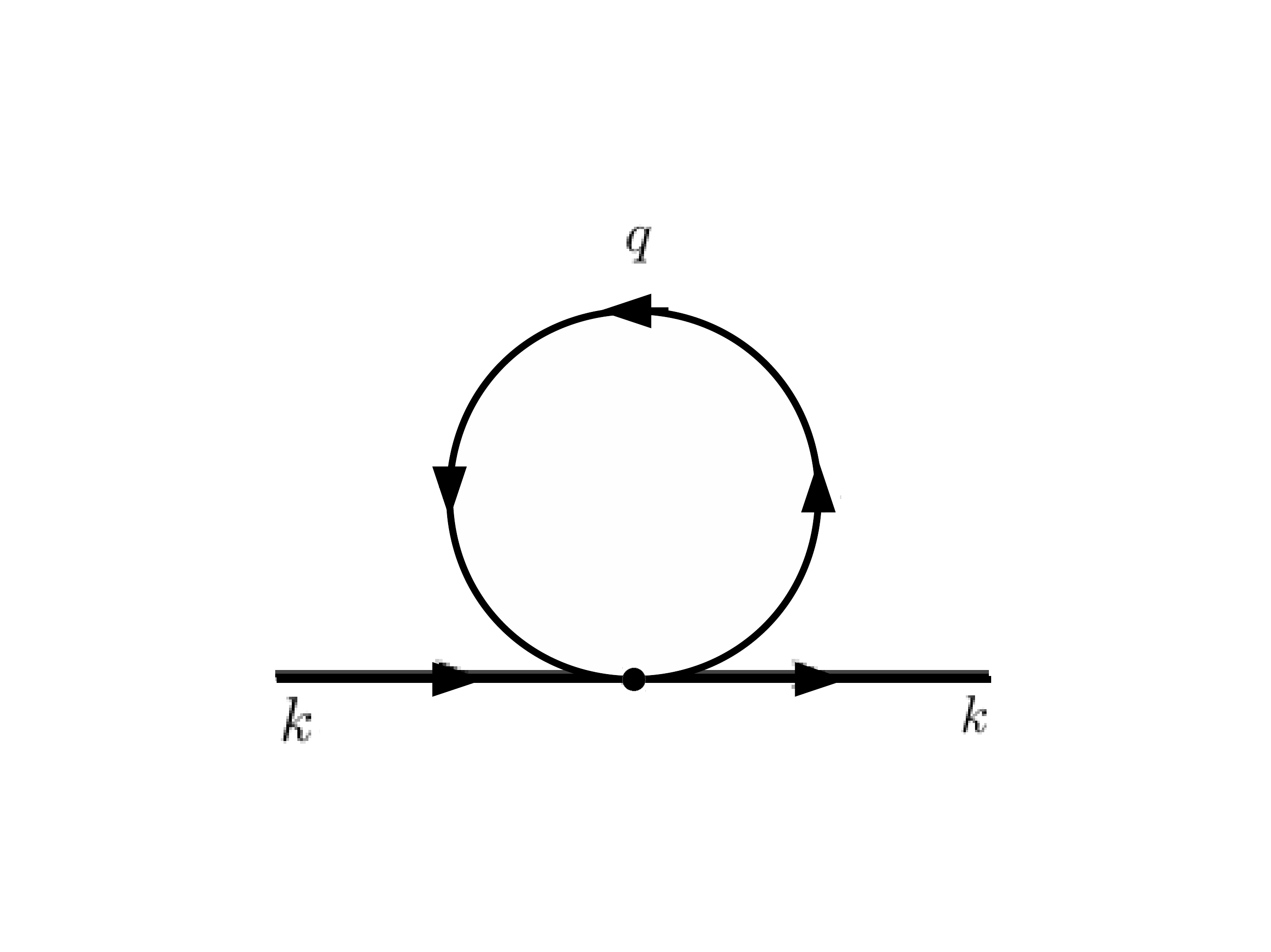}
\caption{One-loop self-energy diagram in the in-plateau case with its local fermion-fermion interaction.}
\label{fig:oneloopSEdeconfined}
\end{figure}

\begin{figure}
\centering
 a)\includegraphics[scale=0.15]{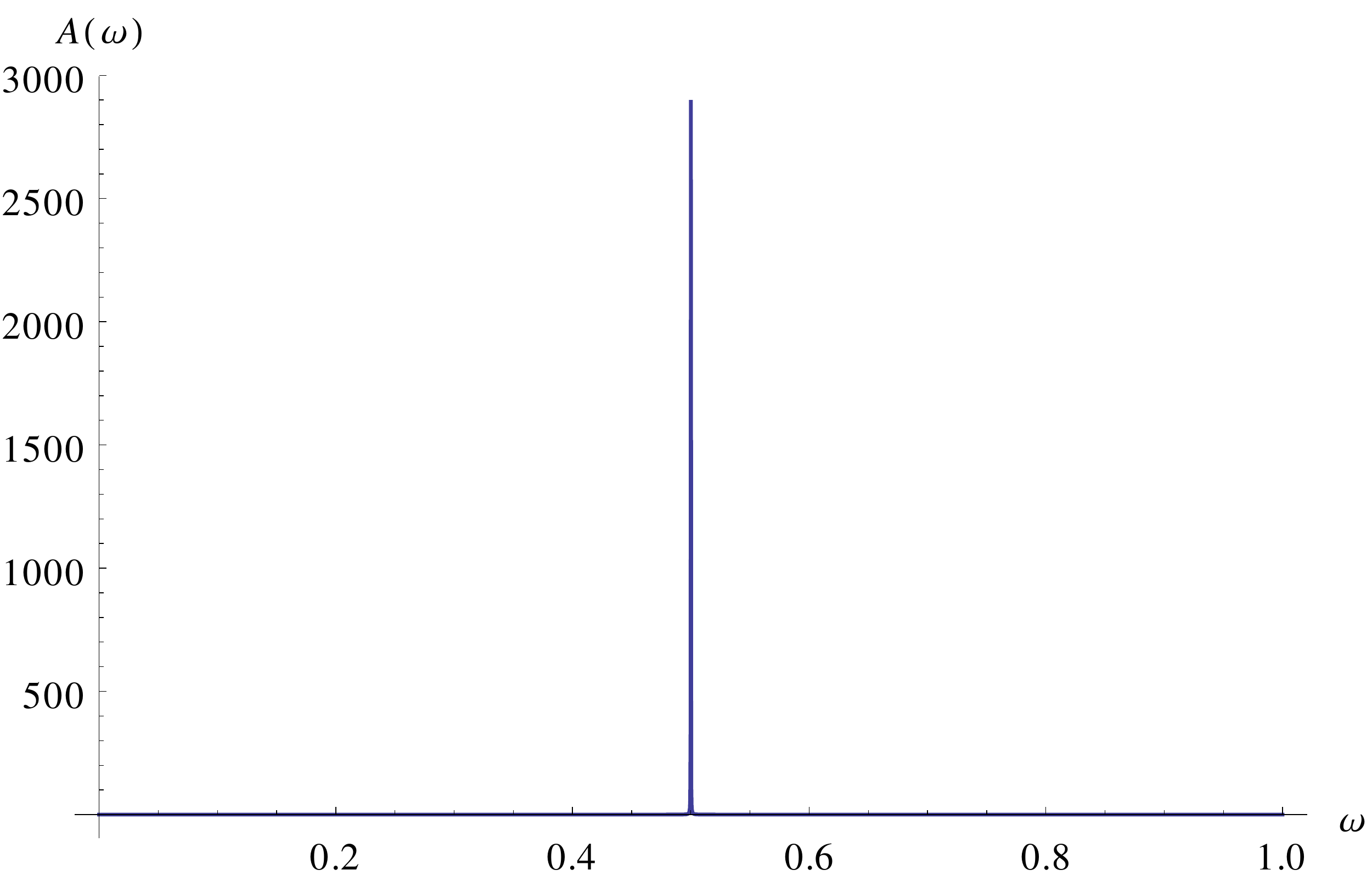}
 b)\includegraphics[scale=0.15]{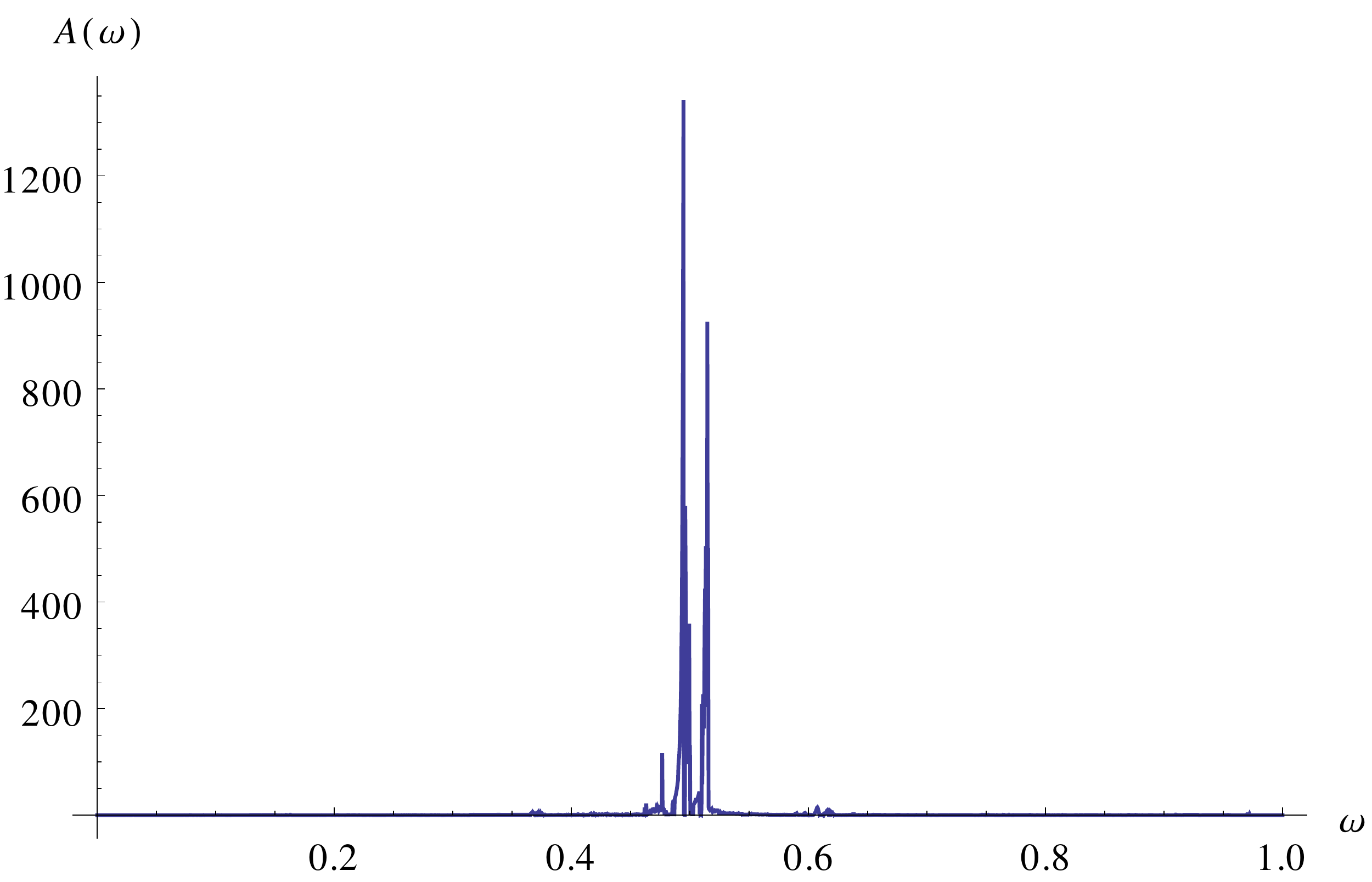}
 \caption{Spectral function $A(\omega)$ at a fixed $|\mathbf{k}|$ for quadratic dispersion around Fermi point a) in the in-plateau case
 b) in the out-of-plateau case \cite{deltapeak}.
  }
\label{fig:SFconfined}
\end{figure}

For the in-plateau case, where we have local interaction, we compute the
one-loop fermion self-energy diagram shown in the Fig.~\ref{fig:oneloopSEdeconfined} with 4-fermion vertex given in Eq. (\ref{kerneldeconf}) 
from which we obtain for the one-loop self-energy $\Sigma(\mathbf{k},\omega)=\int \frac{d^3q}{(2\pi)^3}G_0(q)V(k,q)$
where we have to take into account the fact that there are four equivalent configurations of the Feynman diagram in Fig.~\ref{fig:oneloopSEdeconfined},
contributing to $\Sigma(\mathbf{k},\omega)$.
The resulting spectral function is demonstrated in Fig.~\ref{fig:SFconfined}a).
We observe that with the local (or short-range) interaction of the in-plateau state, the sharp spectral peak of free fermions
is not significantly broadened or dispersed.

\begin{figure}
 \centering
a)\includegraphics[scale=0.140]{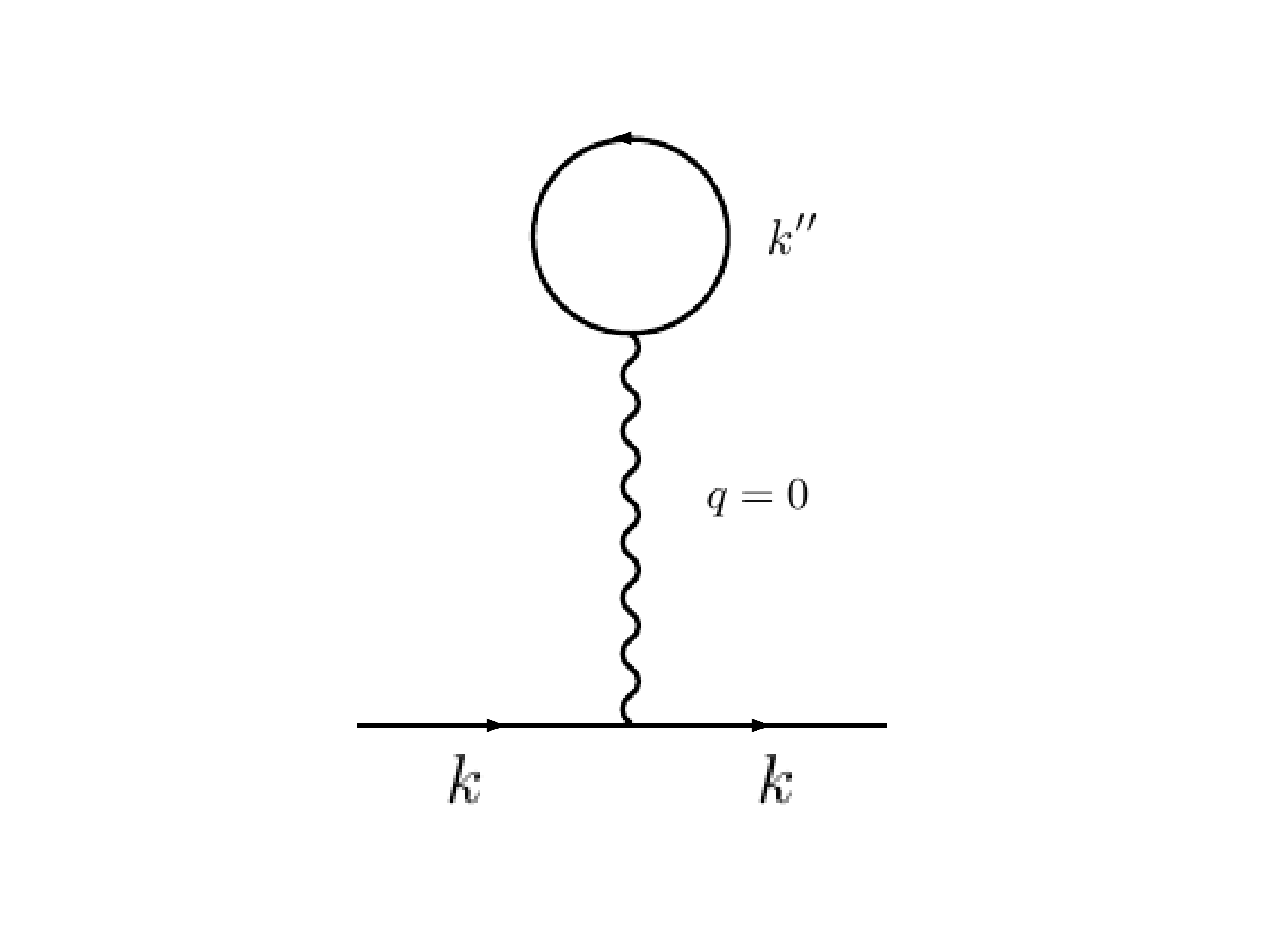}
b)\includegraphics[scale=0.140]{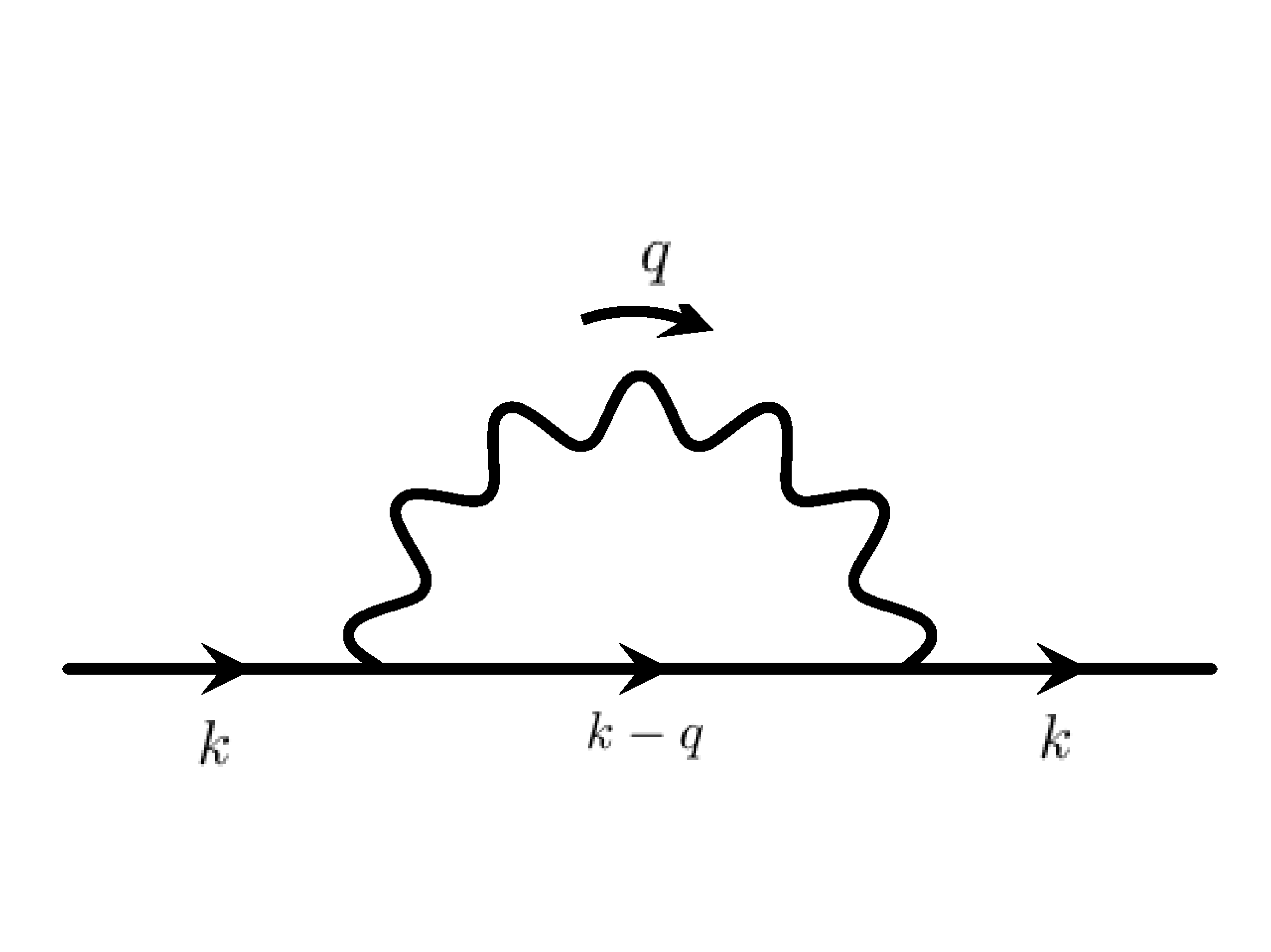}
\caption{One-loop self-energy diagrams in the out-of-plateau case where fermion interaction is long-ranged; a) The tadpole diagram b) The bubble diagram.}
\label{fig:oneloopSEconfined}
\end{figure}
For the out-of-plateau case, the one-loop self-energy diagrams are shown in Fig.~\ref{fig:oneloopSEconfined} \cite{AGD}
where the nonlocal long-range interaction is represented by wiggling (wavy) line. The kernel is given by Eq. (\ref{kernelconf})
with $q=k=k'$ as the momentum-frequency of the spin sector's gauge field  which mediates the long-range interaction, $k'',k''''$ are the momenta-frequencies of the scattered fermions.
From this expression, it is clear that the contribution of tadpole diagram in Fig.~\ref{fig:oneloopSEconfined} (a) vanishes because momentum conservation forces $q=0$. 
The expression for the nonvanishing diagram in Fig.~\ref{fig:oneloopSEconfined}(b) is $\Sigma(k)=\int \frac{d^{3}q}{(2\pi)^{3}}G_0(k-q)V(k,q)$
with $V(q,k)$ given in Eq. (\ref{kernelconf}) and where we should note that there are two equivalent configurations of this diagram with equal contribution.
The self-energy is both momentum and frequency dependent, reflecting the non-instantaneousness of the algebraic long-range interaction.

\begin{figure}
\centering
a)\includegraphics[scale=0.15]{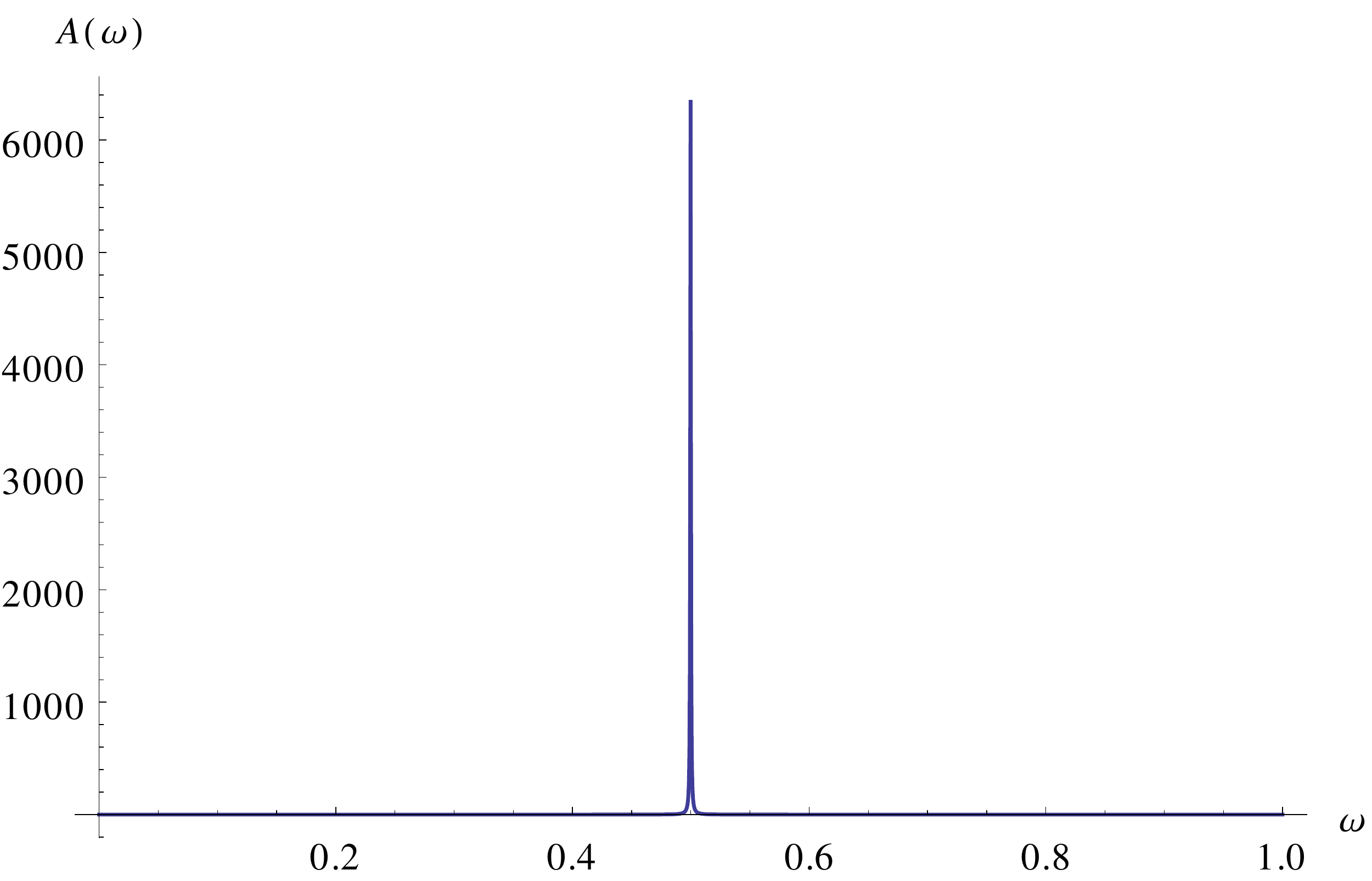}
b)\includegraphics[scale=0.15]{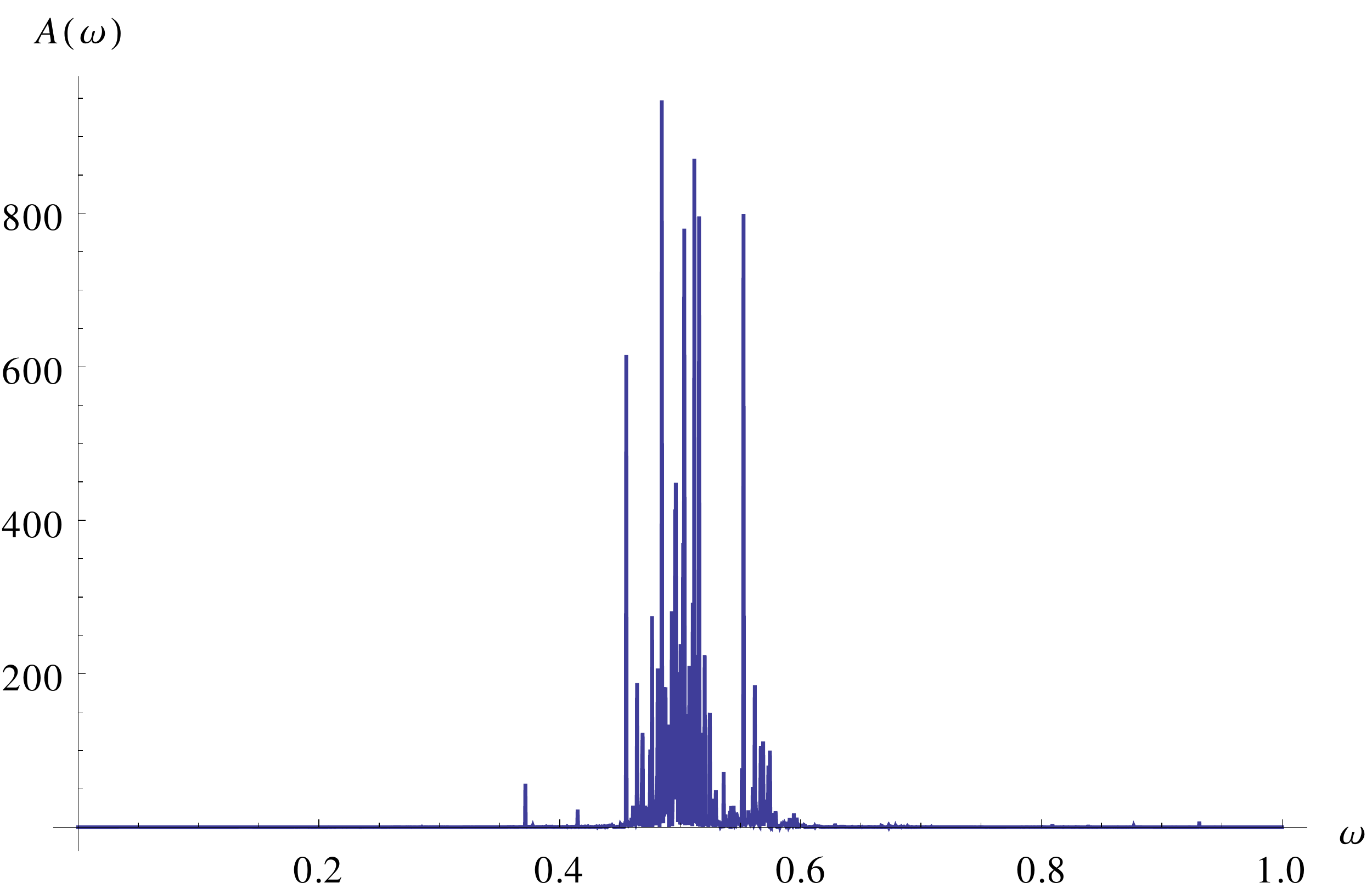}
 \caption{Spectral function $A(\omega)$ at a fixed $|\mathbf{k}|$ for linearized dispersion at finite doping a) in the in-plateau case b) in the out-of-plateau case \cite{deltapeak}.}
\label{fig:lineardispersionSFconfined}
\end{figure}

We show the resulting profile of $A(\omega)$ at a fixed $\mathbf{k}$ in Fig.~\ref{fig:SFconfined}b) for this off-plateau case with quadratic hole dispersion.
We notice that, due to the the algebraically decaying long-range fermion-fermion interaction, the spectral weight is heavily broadened compared to that of free noninteracting fermions 
which has hallmark delta function peak. The spectral peak broadening increases with the strength of the coupling to gauge field represented by gauge charge $e_g$
and also the Goldstone mode's total energy bandwidth $\delta q_0\sim 2v_b \Lambda$, where $2\Lambda$ is the total momentum bandwidth.
In the original microscopic spin model Eq. (\ref{latticespinmodel}), this is achieved for large $J,D\gg h$.
Comparing the two cases, it can be seen that the hole spectral function
in the out-of-plateau state is much more significantly broadened and suppressed compared to that of the in-plateau state.
This broadening reflects the effects of Goldstone bosons which survive outside the plateau and mediate the long-range interaction.
We then consider the more realistic finite hole doping situation with its linear dispersion with results shown in Figs.~\ref{fig:lineardispersionSFconfined}a) and b) giving the same conclusions.

\textit{Discussion.\textemdash}
We have demonstrated that the fermion spectral function of hole-doped antiferromagnet can be used as a direct probe of on-plateau vs. off-plateau physics of the spin sector.
We have shown that within plateau the spin sector generates local fermion-fermion interaction while outside plateau it generates long-range fermion-fermion interaction with both
density-density and dipolar contents.
This difference manifests in the spectral function of the holes.
In particular, our result predicts that the hole spectral function for the in-plateau case remains a sharp delta function hallmark of free fermion
spectral function with negligible broadening, whereas outside plateau, the hole spectral function is significantly broadened and reduced in height, subject to an appropriate sum rule.
We also predict that finite hole doping will shift the magnitude of plateaus.

With the presence of long-range algebraic interactions, there is a possibility for the formation of Wigner crystal \cite{wigner} of holes, 
when the density-density interaction, which is indeed repulsive in this case, dominates over dipolar interaction and kinetic energies.
In contrast to the usual Coulomb case however, based on dimensional analysis, we expect the Wigner crystal to occur at high density of holes rather than low density. 
This is due to the fact that the algebraic interaction decays as $V(r)\sim 1/r^3$ rather than the usual $V(r)\sim 1/r$, with kinetic energy goes as $1/r^2$.

Compared with 1-d case, it is expected that, other than the clear differences in technical details, 
the distinction in the behavior of spectral function
on and off plateaus will be less discernible, due to the Luttinger (non-Fermi) liquid behavior. Our results qualitatively agree with 
hole spectral function theoretical calculations for antiferromagnet in underdoped Cuprates (at zero field) treated with slave-particle approach, where similar broadening arises due to 
the nonlocal (despite finite-ranged few nearest-neighbor) interactions \cite{HoleSF} and confirmed experimentally \cite{Exp}.
As photoemission studies on hole-doped antiferromagnets with plateaus at finite field themselves have not yet been available, we would like to propose candidate materials:
2-d antiferromagnet compounds $\mathrm{SrCu_2(BO_3)_2}$ \cite{2-dAFmaterial} and $\mathrm{(CuBr)Sr_2Nb_3O_{10}}$ \cite{compound} 
which are very promising compounds for testing our theoretical predictions as they are 
2-d antiferromagnetic materials that have been shown to display plateaus.

\textit{Acknowledgements.\textemdash}IM is supported by the grant No. ANR-10-LABX-0037 of the Programme des Investissements d'Avenir of France.
The authors thank M. Oshikawa and P. Romaniello for very helpful and insightful discussions.
PP would like also to thank C. Lamas for many discussions closely related to this subject.

\newpage

\clearpage

\begin{widetext}
\begin{center}

\textbf{Hole Properties In and Out of Magnetization Plateau in 2-d Antiferromagnet}\\

Supplementary Material\\

\bigskip

Imam Makhfudz and Pierre Pujol\\
%
Laboratoire de Physique Th\'{e}orique--IRSAMC, CNRS and Universit\'{e} de Toulouse, UPS, F-31062 Toulouse, France\\ 
\date{\today}
\bigskip

\textbf{Derivation of Fermion Effective Action}
 
\end{center}

\end{widetext}

The effective action for fermion obtained from integrating the spin sector scalar field is derived using functional integral formalism in Euclidean space-time. 
From the full action Eqs. (\ref{spinsectoraction}) and (\ref{holeaction}), we can write the partition function as 

\begin{equation}\label{partitionf}
 Z=\int D\overline{\psi}D\psi e^{-S^0_{\overline{\psi},\psi}}\int D\phi e^{-\int_{k,k'}\phi^*_kG^{-1}(k)\phi_{k'}\delta(k-k')-\int f_k \phi_k}
\end{equation}
First, we consider quadratic dispersion $\epsilon^0_k=t(k^2_x+k^2_y)$ for hole valid near the minimum of the band, 
which will be at Fermi level and forms Fermi point when chemical potential $\mu=-4t$ corresponding to zero or thermodynamically small hole doping. 
That is, we can write noninteracting kinetic fermionic hole action
\begin{equation}
 S^0_{\overline{\psi},\psi}=\int_r \overline{\psi}(r)(\partial_{\tau}-t\nabla^2)\psi(r)=\int_k \overline{\psi}_k (ik_0+t\mathbf{k}^2)\psi_k
\end{equation}
The inverse kernel in Eq. (\ref{partitionf}) is given by  
\begin{equation}
  G^{-1}(k)=[\frac{K_{\tau}}{2}k^2_0 + \frac{K_{r}}{2}\mathbf{k}^2+\frac{1}{2}m^2_{\phi}]
\end{equation}
where the last term (effective mass) will later determine whether one is in plateau or out of plateau.
The hole is coupled to gauge field from spin sector which can be represented by minimal coupling $k_{\mu}\rightarrow k_{\mu}-eA_{\mu}$.
In the whole following derivation, we will set the boson velocity to unity $v_b=1$ for brevity.
We obtain action of quadratic dispersed fermion coupled to gauge field

\begin{equation}
 S_{\overline{\psi},\psi}= \int_{k's} \overline{\psi}_{k'''} (ik''_0+e_gk'_0\phi_{k'}+t(\mathbf{k}''-ie_g\mathbf{k}'\phi_{k'})^2)\psi_{k''}\delta(\sum_{k's})
\end{equation}
The mass of the vortex loops receives correction from the fermion-gauge field coupling
\begin{widetext}
\begin{equation}\label{mass}
  \frac{1}{2}m^2_{\phi}(k,k')=\frac{1}{2}m^2_{\phi}\delta(k'-k)+te^2_g\int_{k'',k'''}\mathbf{k}^2\overline{\psi}_{k'''}\psi_{k''}\delta(-k'''+k''+k'-k)
\end{equation}
\end{widetext}
The correction (second) term in Eq. (\ref{mass})\comment{comes from the coupling of the spin sector to the hole which and explicitly contains fermion bilinear operator but this term} 
vanishes as $|\mathbf{k}|\rightarrow 0$, 
which should indeed be the case since we assumed classical ground state with broken global continuous symmetry and the associated massless Nambu-Goldstone modes.
In the' zeroth order approximation',\comment{as the second term is of order $\mathcal{O}(e^2_g)$,} we can take $m^2_{\phi}(k,k')=m^2_{\phi}\delta(k-k')$.
This is justified in the low energy limit $|\mathbf{k}|\rightarrow 0$ and by the observation that the correction term is of order $\alpha\sim e^2_g$, which is the small
parameter in the perturbation expansion we are doing.

The function $f_k$ multiplying linear term in Eq. (\ref{partitionf}) is 
\begin{widetext}
\begin{equation}\label{linearfunction}
 f_k=-(\frac{S-m}{a^2})k_0 -2ite_g\int_{k'',k'''}\mathbf{k}\cdot\mathbf{k}''\overline{\psi}_{k'''}\psi_{k''}\delta(-k'''+k''+k)+\int_{k',k''}\overline{\psi}_{k'''}e_gk_0\psi_{k''}\delta(k+k''-k''')
\end{equation}
\end{widetext}
where in this derivation, $k,k'$ are reserved for the momentum-frequency of the spin sector scalar field $\phi$ while the other $k$'s are the momenta-frequencies of the fermions.
Integrating out the bosonic scalar field $\phi$ in Eq.(\ref{partitionf}), we have
\[
 Z=\int D\overline{\psi}D\psi e^{-\int_k\overline{\psi}_k\epsilon^0_k\psi_k-\frac{1}{4}\int_{k,k'}f^*_kG(k)f_{k'}\delta(k-k')}
\]
\begin{equation}
 =\int D\overline{\psi}D\psi e^{-(S^0_{\overline{\psi},\psi}+\delta S^{\mathrm{quadratic}}_{\overline{\psi},\psi})}
\end{equation}
\comment{where $\epsilon^0_k=t(k^2_x+k^2_y)$ for $\mu=-4t$.} The first term in Eq. (\ref{linearfunction}), coming from Berry phase in Eq. (\ref{spinsectoraction}), gives rise to constant energy shift 
plus small ($\mathcal{O}(\sqrt{\alpha})\sim e_g$) correction to bilinear fermion action,
\begin{equation}
 \delta S^{\mathrm{kinetic}}_{\overline{\psi}\psi}=-2(\frac{S-m}{a^2})e_g\int_{k's}k'^2_0G(k')\overline{\psi}_{k'''}\psi_{k''}\delta(\sum_{k's})
\end{equation}
but we are more interested in fermion-fermion interaction. Considering the zeroth order approximation 
\comment{in the expansion of $\overline{\psi}\psi$ in $G^{-1}=-\partial^2+m^2_{\phi}/2$} mentioned previously, we obtain

\begin{widetext}
\comment{
\[
 \delta S^{\mathrm{quadratic}}_{\overline{\psi},\psi}=\frac{1}{4}\int_{k's}\overline{\psi}_{k'''}\psi_{k''}\left(e_gk_0+2ite_g\mathbf{k}\cdot\mathbf{k}'''\right)G(k-k')\left(e_gk'_0-2ite_g\mathbf{k}'\cdot\mathbf{k}'''''\right)\overline{\psi}_{k'''''}\psi_{k''''}\delta(\sum_{k's})
\]}
\begin{equation}\label{holeactionrenormalizationKspacesupp}
\delta S^{\mathrm{quadratic}}_{\overline{\psi},\psi}=\int_{k's}\overline{\psi}_{k'''}\psi_{k''} \frac{e^2_g[\frac{1}{4}k^2_0+t^2(\mathbf{k}\cdot\mathbf{k}'')(\mathbf{k}\cdot\mathbf{k}'''')]}{\frac{K_{\tau}}{2}k^2_0 + \frac{K_{r}}{2}\mathbf{k}^2+\frac{1}{2}m^2_{\phi}} \overline{\psi}_{k'''''}\psi_{k''''}\delta(\sum_{k's})
\end{equation}
\end{widetext}
as given in the main text, with superscript '$\mathrm{quadratic}$' refers to quadratic dispersion (to be abbreviated as 'quad' whenever necessary). We have taken into account the fact that $k$ and $k'$ are both
momentum-frequency of the gauge field and therefore, eventually $k=k'$.\comment{ We have set aside 'cross terms' and focus on the two terms given in Eq. (\ref{holeactionrenormalizationKspace}).}
\comment{We note that the overall interaction is repulsive when $\mathbf{k}''',\mathbf{k}'''''$ 
of the scattered fermions
are both parallel (or have component parallel) to the gauge field momentum $\mathbf{k},\mathbf{k}'=\mathbf{q}$. Also, the overall strength of this interaction is proportional to $t^2e^2_g/K_s$. 
Kinetic energy on the other hand has overall
strength proportional to $t$. We can then define a dimensionless ratio $\eta=te^2_g/K_s\sim tS^2(4J+D)^2/(Jh^2)$ giving the relative magnitude of potential over kinetic energies.}
\comment{The resulting fermion-fermion interaction contains density-density interaction\comment{from $e_gk_0e_gk'_0$} term and a term which is not of density-density type.}
\comment{We will retain only two most important terms.}\comment{This results in kernel
\begin{equation}
V^{\mathrm{out-of-plat}}(k's)=\frac{2}{K_s}e^2_g\frac{[t^2(\mathbf{q}\cdot\mathbf{k}''')(\mathbf{q}\cdot\mathbf{k}''''')+\frac{1}{4}q^2_0]}{q^2_0 + \mathbf{q}^2+\tilde{m}^2_{s}}
\end{equation}}

The real space form of the above 4-fermion interaction is found to be

\begin{widetext}
\begin{equation}\label{holeactionrenormalizationXspace}
\delta S^{\mathrm{quad}}_{\overline{\psi},\psi}=\frac{e^2_g}{2K_s}\int_{x_{\mu},x'_{\mu'}}\rho(x)\left(\partial_{\tau}\partial_{\tau'}\frac{e^{-\tilde{m}_s|x'_{\mu}-x_{\mu}|}}{4\pi|x'_{\mu}-x_{\mu}|}\right)\rho(x')
+
\frac{2t^2e^2_g}{K_s}\int_{x_{\mu},x'_{\mu'}}\left(\overline{\psi}(x)[\nabla_{\mathbf{x}}\psi(x)]\cdot\nabla_{\mathbf{x}}\right)\left(-\nabla_{\mathbf{x}'}\frac{e^{-\tilde{m}_s|x'_{\mu}-x_{\mu}|}}{4\pi|x'_{\mu}-x_{\mu}|}\overline{\psi}(x')\right)\cdot[\nabla_{\mathbf{x}'}\psi (x')]
\end{equation}
\end{widetext}
\comment{plus 'cross terms', in addition to the}which takes the form of density-density and dipole-dipole (to be explained below) interactions. 
Here, $\rho(x)=\overline{\psi}(x)\psi(x)$ is fermion density operator, 
$x_{\mu}=(\tau,\mathbf{x})$ is Euclidean space-time coordinate,$|x'_{\mu}-x_{\mu}|=\sqrt{(x'-x)^2+(y'-y)^2+v^2_b(\tau'-\tau)^2}$, and the spatial derivatives 
$\nabla_x,\nabla_{x'}$ in the middle act only on the kernel. We have set $K_{\tau}=K_r=K_s$ in obtaining Eq. (\ref{holeactionrenormalizationXspace}), which also means we 
set the boson velocity to unity $v_b=1$ as mentioned earlier\comment{in this derivation, for brevity}.

In the case of confined vortex loops out-of-plateau, we evaluate twice derivatives of the kernel and take $m_{\phi}\rightarrow 0$ in the end, giving us

\begin{widetext}
\[
\delta S^{\mathrm{quadratic-out-of-plateau}}_{\overline{\psi},\psi}=-\frac{1}{2K_s}e^2_g\int_{x_{\mu},x'_{\mu'}}\rho(x)\left(\frac{3(\tau'-\tau)^2-|x'_{\mu}-x_{\mu}|^2}{4\pi|x'_{\mu}-x_{\mu}|^5}\right)\rho(x')
\]
\[
+t^2e^2_g \frac{2}{ K_s}\int_{x_{\mu},x'_{\mu'}}\overline{\psi}(x)[\partial_{x}\psi(x)]\frac{3(x'-x)^2-|x'_{\mu}-x_{\mu}|^2}{4\pi|x'_{\mu}-x_{\mu}|^5}\overline{\psi}(x')[\partial_{x'}\psi(x')]+\overline{\psi}(x)[\partial_{y}\psi(x)]\frac{3(y'-y)^2-|x'_{\mu}-x_{\mu}|^2}{4\pi|x'_{\mu}-x_{\mu}|^5}\overline{\psi}(x')[\partial_{y'}\psi(x')]
\]
\begin{equation}\label{holeactionrenormalizationCONFINED}
+\frac{3(x'-x)(y'-y)}{4\pi|x'_{\mu}-x_{\mu}|^5}\left(\overline{\psi}(x)[\partial_x\psi(x)]\overline{\psi}(x')[\partial_{y'}\psi(x')]+\overline{\psi}(x)[\partial_y\psi(x)]\overline{\psi}(x')[\partial_{x'}\psi(x')]\right)
\end{equation}
\end{widetext}
We note that we obtain Coulombic-like algebraically decaying long-range density-density interaction in the first term and dipolar interaction in the 
remaining terms\comment{(we stress these two terms and put aside 'mixed terms')}.
The dipolar nature of the remaining terms is indicated by the presence of derivatives on fermion fields and the form of the kernel. 
This is nontrivial 4-fermion interaction but can be treated with field theoretical perturbation theory \cite{AGDsuppmat}.

In the in-plateau deconfined $m_{\phi}\rightarrow \infty$ limit on the other hand, we can make use of the fact that the kernel behaves as rapidly decaying function and so we can approximate it as Dirac delta function, 
with appropriately determined amplitude.
To be precise, $\exp(-\tilde{m}_s r)/(4\pi r)\simeq A\delta^{(3)}(r)$ with $A=1/\tilde{m}^2_{s}$ in (2+1)-D. This gives us

\begin{widetext}
\[
\delta S^{\mathrm{quadratic-in-plateau}}_{\overline{\psi},\psi}=\frac{1}{2K_s}e^2_g\int_{x_{\mu},x'_{\mu'}}\rho(x)(\partial_{\tau}\partial_{\tau'}A\delta^{(3)}(x'_{\mu}-x_{\mu}))\rho(x')
\]
\[
+t^2e^2_g \frac{2}{K_s}\int_{x_{\mu},x'_{\mu'}}\left(\overline{\psi}(x)[\nabla_{x}\psi(x)]\cdot\nabla_{x}\right)\left(-\nabla_{x'}A\delta^{(3)}(x'_{\mu}-x_{\mu})\cdot\overline{\psi}(x')[\nabla_{x'}\right)\psi (x')]
\]
\[
=\frac{1}{2K_s}Ae^2_g\int_{x_{\mu}}\left(\partial_{\tau}\rho(x)\right)^2-At^2e^2_g \frac{2}{K_s}\int_{x_{\mu}}(\partial_x[\overline{\psi}(x)[\partial_{x}\psi(x)]]\partial_x[\overline{\psi}(x)\partial_{x}\psi(x)]]+\partial_y[\overline{\psi}(x)[\partial_{y}\psi(x)]]\partial_y[\overline{\psi}(x)[\partial_{y}\psi(x)]]
\]
\begin{equation}\label{holeactionrenormalizationDECONFINED1}
+2\partial_x[\overline{\psi}(x)[\partial_{x}\psi(x)]]\partial_y[\overline{\psi}(x)[\partial_{y}\psi(x)]])
\end{equation}
\[
=Ae^2_g \frac{2}{K_s}\int_{k,k',k'',k'''}[-\frac{1}{4} (k'''''_0-k''''_0)(k'''_0-k''_0)-t^2((k''^2_x-k'''_xk''_x)(k''''^2_x-k'''''_xk''''_x)+(k''^2_y-k'''_yk''_y)(k''''^2_y-k'''''_yk''''_y)
\]
\begin{equation}\label{holeactionrenormalizationDECONFINED2}
+(k''^2_x-k'''_xk''_x)(k''''^2_y-k'''''_yk''''_y)+(k''^2_y-k'''_yk''_y)(k''''^2_x-k'''''_xk''''_x))]\overline{\psi}_{k'''}\psi_{k''}\overline{\psi}_{k'''''}\psi_{k''''}\delta(k''''-k'''''+k''-k''')
\end{equation}
\end{widetext}
where we obtain a rather delicate but otherwise local 4-fermion interaction. The resulting net kernel is 
\[
  V^{\mathrm{in-plat}}(k's)=\frac{2}{K_s}Ae^2_g[-t^2(\mathbf{k}'''\cdot\mathbf{k}''-\mathbf{k}''^2)(\mathbf{k}'''''\cdot\mathbf{k}''''-\mathbf{k}''''^2)
\]
\begin{equation}\label{kerneldeconfapp}
  -\frac{1}{4}(k'''''_0-k''''_0)(k'''_0-k''_0)]
\end{equation}
as given in Eq.(\ref{kerneldeconf}) in the main text.

Now, we present some details on the derivation of effective 4-fermion interaction term for linear dispersion case, applicable at finite doping.
We start from fermion bilinear action with linear dispersion minimally coupled to gauge field, given in Eq. (\ref{linearfermionkineticaction}). 
\begin{widetext}
\begin{equation}\label{linearfermionkineticaction}
 S_{\overline{\psi},\psi}=\int_{k's}\overline{\psi}_{k'''}[ik''_0+e_gk'_0\phi_{k'}+2t[(k''_x-k''_{xF}-ie_gk'_x\phi_{k'})\sin k''_{xF}+(k''_y-k''_{yF}-ie_gk'_y\phi_{k'})\sin k''_{yF}]]\psi_{k''}\delta(\sum_{k's})
\end{equation}
\end{widetext}
where $k_{xF}=k_F \cos\theta,k_{yF}=k_F\sin\theta$, with $\theta$ is the polar angle along the nearly circular Fermi surface defined by $k^2_{xF}+k^2_{yF}=k^2_F=4+\mu/t$ 
and as before, we have used $eA_{\mu}=e_g\partial_{\mu}\phi$.
Integrating out the gauge field, we obtain 

\begin{widetext}
\begin{equation}\label{lineardispersionregime4fermionapp}
 \delta S^{\mathrm{linear}}_{\overline{\psi},\psi}=\frac{1}{4}e^2_g\int_{k's}\overline{\psi}_{k'''}\psi_{k''}\left(\frac{k_0k'_0}{\frac{K_{\tau}}{2}k^2_0 + \frac{K_{r}}{2}\mathbf{k}^2+\frac{1}{2}m^2_{\phi}}+ 4t^2k^2_F\frac{\mathbf{k}\cdot\mathbf{k}''}{|\mathbf{k}''|}\frac{1}{\frac{K_{\tau}}{2}k^2_0 + \frac{K_{r}}{2}\mathbf{k}^2+\frac{1}{2}m^2_{\phi}}\frac{\mathbf{k}'\cdot\mathbf{k}''''}{|\mathbf{k}''''|}\right)\overline{\psi}_{k'''''}\psi_{k''''}\delta(\sum_{k's})
\end{equation}
\end{widetext}
where we have made use of approximation $\sin k_{xF}\approx k_{xF},\sin k_{yF}\approx k_{yF}$ valid at finite but low doping levels, where $k_F\lesssim 1$. This linearized fermion action has equivalent momentum dependence
to that of 1-d theory with left and right mover fermions with linear dispersion. 
\comment{
which, by power counting of momenta-frequency, should give the same qualitative result as the 1-d case, except that we now have a whole spectrum of fermions along the Fermi surface (not just left and right movers) and we 
are not able to bosonize the problem.}
The $1/|\mathbf{k}''|,1/|\mathbf{k}''''|$ factors however, pose technical difficulty as their inverse Fourier transforms are not well defined. In this case, $\mathbf{k}'',\mathbf{k}''''$ are the momenta
of fermions. To handle this, we approximate the integral over the whole Fourier space with integral over Fermi surface, for which $|\mathbf{k}''|=|\mathbf{k}''''|=k_F$. 
With this, we have Eq. (\ref{lineardispersionregime4fermionapp}) as the final form of effective fermion action,

\begin{widetext}
\begin{equation}\label{lineardispersionregime4fermionFINAL}
 \delta S^{\mathrm{linear}}_{\overline{\psi},\psi}=\frac{1}{4}e^2_g\int_{k's}\overline{\psi}_{k'''}\psi_{k''}\left(\frac{k_0k'_0}{\frac{K_{\tau}}{2}k^2_0 + \frac{K_{r}}{2}\mathbf{k}^2+\frac{1}{2}m^2_{\phi}}+ 4t^2(\mathbf{k}\cdot\mathbf{k}'')\frac{1}{\frac{K_{\tau}}{2}k^2_0 + \frac{K_{r}}{2}\mathbf{k}^2+\frac{1}{2}m^2_{\phi}}(\mathbf{k}'\cdot\mathbf{k}'''')\right)\overline{\psi}_{k'''''}\psi_{k''''}\delta(\sum_{k's})
\end{equation}
\end{widetext}
We have verified that using approximate linear cone dispersion $\epsilon_k=v_F(|\mathbf{k}|-k_F)$ gives rise to, remarkably, precisely the same expression for effective action 
as the one obtained above using linearized dispersion
$\epsilon_k=2t[(k_x-k_{xF})\sin k_{xF}+(k_y-k_{yF})\sin k_{yF}]$.

Again, considering the confined $m_{\phi}\rightarrow 0$ limit out-of-plateau, we obtain

\begin{widetext}
\[
\delta S^{\mathrm{linear-out-of-plateau}}_{\overline{\psi},\psi}=-\frac{1}{2K_s}e^2_g\int_{x_{\mu},x'_{\mu'}}\rho(x)(\frac{3(\tau'-\tau)^2-|x'_{\mu}-x_{\mu}|^2}{4\pi|x'_{\mu}-x_{\mu}|^5})\rho(x')
\]
\[
+t^2e^2_g \frac{2}{K_s}\int_{x_{\mu},x'_{\mu'}}\overline{\psi}(x)[\partial_{x}\psi(x)]\frac{3(x'-x)^2-|x'_{\mu}-x_{\mu}|^2}{4\pi|x'_{\mu}-x_{\mu}|^5}\overline{\psi}(x')[\partial_{x'}\psi(x')]+\overline{\psi}(x)[\partial_y\psi(x)]\frac{3(y'-y)^2-|x'_{\mu}-x_{\mu}|^2}{4\pi|x'_{\mu}-x_{\mu}|^5}[\partial_{y'}\overline{\psi}(x')]\psi(x')
\]
\begin{equation}\label{holeactionrenormalizationCONFINED}
+\frac{3(x'-x)(y'-y)}{4\pi|x'_{\mu}-x_{\mu}|^5}\left(\overline{\psi}(x)[\partial_x\psi(x)]\overline{\psi}(x')[\partial_{y'}\psi(x')]+\overline{\psi}(x)[\partial_{y}\psi(x)]\overline{\psi}(x')[\partial_{x'}\psi(x')]\right)
\end{equation}
\end{widetext}
where the relevant fermions contributing to the integral are implicitly constrained to live near the Fermi surface.

In the deconfined limit $m_{\phi}\rightarrow \infty$ in the plateau, we should get 

\begin{widetext}
\[
\delta S^{\mathrm{linear-in-plateau}}_{\overline{\psi},\psi}=\frac{1}{2K_s}e^2_g\int_{x_{\mu},x'_{\mu'}}\rho(x)(\partial_{\tau}\partial_{\tau'}A\delta^{(3)}(x'_{\mu}-x_{\mu}))\rho(x')
\]
\[
+t^2e^2_g \frac{2}{K_s}\int_{x_{\mu},x'_{\mu'}}\left(\overline{\psi}(x)[\nabla_{x}\psi(x)]\cdot\nabla_{x}\right)\left(-\nabla_{x'}A\delta^2(\mathbf{x}'-\mathbf{x})\overline{\psi}(x')\right)\cdot[\nabla_{x'}\psi (x')]
\]
\[
=\frac{1}{2K_s}Ae^2_g\int_{x_{\mu}}\left(\partial_{\tau}\rho(x)\right)^2-At^2e^2_g \frac{2}{K_s}\int_{x_{\mu}}(\partial_x[\overline{\psi}(x)[\partial_{x}\psi(x)]]\partial_x[\overline{\psi}(x)[\partial_{x}\psi(x)]]+\partial_y[\overline{\psi}(x)[\partial_{y}\psi(x)]]\partial_y[\overline{\psi}(x)[\partial_{y}\psi(x)]]
\]
\begin{equation}\label{linearholeactionrenormalizationDECONFINED1}
+\partial_x[\overline{\psi}(x)[\partial_{x}\psi(x)]]\partial_y[\overline{\psi}(x)[\partial_{y}\psi(x)]]+\partial_y[\overline{\psi}(x)[\partial_{y}\psi(x)]]\partial_x[\overline{\psi}(x)[\partial_{x}\psi(x)]])
\end{equation}
\[
=Ae^2_g \frac{2}{K_s}\int_{k's\approx k_F}[-t^2(k'''_xk''_x-k''^2_x)(k'''''_xk''''_x-k''''^2_x)+(k'''_yk''_y-k''^2_y)(k'''''_yk''''_y-k''''^2_y)+(k'''_xk''_x-k''^2_x)(k'''''_yk''''_y-k''''^2_y)
\]
\begin{equation}\label{linearholeactionrenormalizationDECONFINED2}
+(k'''_yk''_y-k''^2_y)(k'''''_xk''''_x-k''''^2_x))-\frac{1}{4}(k'''''_0-k''''_0)(k'''_0-k''_0)]\overline{\psi}_{k'''}\psi_{k''}\overline{\psi}_{k'''''}\psi_{k''''}\delta(\sum_{k's})
\end{equation}
\end{widetext}
The resulting net kernel is 
\[
  V^{\mathrm{in-plat}}(k's)=\frac{2}{K_s}Ae^2_g[-t^2(\mathbf{k}'''\cdot\mathbf{k}''-\mathbf{k}''^2)(\mathbf{k}'''''\cdot\mathbf{k}''''-\mathbf{k}''''^2)
\]
\begin{equation}\label{kerneldeconfsupp}
  -\frac{1}{4}(k'''''_0-k''''_0)(k'''_0-k''_0)]
\end{equation}
This 4-fermion effective action for linear dispersion on roughly circular Fermi surface is almost the same as that for quadratic dispersion,\comment{ but the details are slightly different.
The derivative operators act on different fermion field operators and}. The sole difference being that the linear dispersion 4-fermion action involves fermions close to the circular Fermi surface whereas the 
quadratic one involves fermions near a Fermi point. The two cases are however smoothly connected as one slowly increases the chemical potential from $\mu=-4t$ up.
We have rechecked the above calculations using QFT's perturbation theory and obtained the same results.


\begin{thebibliography}{1}

\bibitem{DopingMott}P. A. Lee, N. Nagaosa, and X-G. Wen, Rev. Mod. Phys. 78, 17 (2006). 

\bibitem{Natcomm}S. Nishimoto, N. Shibata, and C. Hotta, Nat. Comms. 4,2287 (2013).

\bibitem{plateauexactsolvablemodel}Plateus in hole-doped exactly solvable model however was considered in H. Frahm and C. Sobiella, Phys. Rev. Lett. 83, 5579 (1999).

\bibitem{TTH}A. Tanaka, K. Totsuka, and X. Hu, Phys. Rev. B 79, 064412 (2009). 

\bibitem{LMS}E. H. Lieb, T. D. Schultz, and D. C. Mattis, Ann. Phys. (NY) 16, 407 (1961).

\bibitem{AffLieb}I. Affleck and E. Lieb, Lett. Math. Phys. 12, 57 (1986).

\bibitem{LeeFisher}D.-H. Lee and M. P. A. Fisher, Int. J. Mod. Phys. B 5, 2675 (1991).

\bibitem{TotsukaPLA}K. Totsuka, Phys. Lett. A 228, 103 (1997).

\bibitem{OshikawaAffleck}M. Oshikawa, M. Yamanaka, and I. Affleck, Phys. Rev. Lett. 78, 1984 (1997).

\bibitem{boson1}G. Roux, E. Orignac, P. Pujol, and D. Poilblanc, Phys. Rev. B 75, 245119 (2007). 

\bibitem{boson2}D.C. Cabra, A. De Martino, P. Pujol, and P. Simon, Europhys. Lett.  57, 402 (2002).

\bibitem{boson3}D.C. Cabra, A. De Martino, A. Honecker, P. Pujol, and P. Simon, Phys. Rev. B 63, 094406 (2001).

\bibitem{boson4}D.C. Cabra, A. De Martino, A. Honecker, P. Pujol, and P. Simon, Phys. Lett. A 268, 418 (2000).

\bibitem{Shankar}R. Shankar, Phys. Rev. Lett. 63, 203 (1989); Nuc. Phys. B 330, 433 (1990).

\bibitem{CL-SC-PP}C. A. Lamas, S. Capponi, and P. Pujol, Phys. Rev. B 84, 115125 (2011).

\bibitem{stiffness}Starting from spin model with easy plane anisotropy under magnetic field \cite{TTH}
as given in Eq. (\ref{latticespinmodel}), it can be shown that for 2-d antiferromagnet on square lattice with lattice spacing $a$, 

\begin{equation}
 K_{\tau}=\frac{1}{2a^2(4J+D)},K_r=J(S^2-m^2)
\end{equation}

\bibitem{notation}In the rest of this paper, $k's$ represents the set of all momenta-frequencies appearing in the expression; $k's=k,k',k'',\cdots$,
$\int_{k's}=\int d^3 k/(2\pi)^3\int d^3 k/(2\pi)^3\int d^3 k/(2\pi)^3\cdots$ where $d^3k = dk_0 d^2\mathbf{k}$, and $\delta(\sum_{k's})=\delta(k-k'+k''\cdots)$
imposing the conservation of momentum-frequency.

\bibitem{gaugecoupling}In this case, the gauge coupling $e_{gx}=e_{gy}\sim S(S-m)/m$ while $e_{g\tau}\sim g_2$ where 

\begin{equation}
 g_2=[\frac{1}{2S}(\frac{S-m}{a^2})+\frac{2tSm^{2S-1}\cos k_Fa}{(1-\frac{\delta}{2S})(4J+D)}]
\end{equation} 
at doping level $\delta$. Lorentz invariant theory requires $e_{g\tau}=e_{gx}=e_{gy}=e_g$ which can be achieved by appropriate rescaling of space-time.

\bibitem{S=3per2}T. Sakai and M. Takahashi, Phys. Rev. B 57, R3201 (1998).

\bibitem{circFSdisp}
This linearized dispersion can be approximated by $\epsilon_k=v_F(|\mathbf{k}|-k_F)$ 
with uniform Fermi velocity $v_F=2t \sin k_F$.

\bibitem{SuppMat}Please see the Supplementary Material.

\bibitem{Coulomb}
In this work, we define Coulomb interaction to be that derived from Gauss law $\nabla.\mathbf{E}=-\nabla^2V=\rho/\epsilon_0$, giving, for particles of charge $e$
\[
 H=\int_{\mathbf{k}_1,\mathbf{k}_2,\mathbf{q}}\overline{\psi}_{\mathbf{k}_2-\mathbf{q}}\psi_{\mathbf{k}_2}\frac{4\pi e^2}{q^2}\overline{\psi}_{\mathbf{k}_1+\mathbf{q}}\psi_{\mathbf{k}_1}
\]
The (true) long-rangeness is signalled by the divergence of the kernel $V(q)=\frac{4\pi e^2}{q^2}\rightarrow \infty$ as $q\rightarrow 0$.
Weaker divergence, e.g. $V(q)\rightarrow V(0)$ with $0<V(0)<\infty$ as $q\rightarrow 0$ indicates faster decaying long-range interaction.

\bibitem{PS-QFT}M. Peskin and D. Schroeder, Introduction to Quantum Field Theory (Perseus, Cambridge, MA, 1995).

\bibitem{AGD}A. A. Abrikosov, L. P. Gorkov, and I. E. Dzyaloshinskii, Methods of Quantum Field Theory in Statistical Physics (Dover Publications, 1963).

\bibitem{deltapeak}The plots are given in arbitrary (unspecified) units as only qualitative features are emphasized. 
Conclusions are independent of units or parameters.

\bibitem{wigner}E. Wigner, Phys. Rev. 46, 1002 (1934).

\bibitem{HoleSF}C.L. Kane, P.A. Lee, and N. Read, Phys. Rev. B 39, 6880 (1989),
S.A. Trugman, Phys. Rev. B 41, 892(R) (1990), F. Marsiglio, A. E. Ruckenstein, S. Schmitt-Rink, and C. M. Varma, Phys. Rev. B 43, 10882 (1991).
Broadening also in the sense of appearance of multiple extra subpeaks in the spectral function which redistributes the spectral weight of free fermion or fermion with only local interactions.

\bibitem{Exp}A. Damascelli, Z. Hussain, and Z. Shen, Rev. Mod. Phys. 75, 473 (2003).

\bibitem{2-dAFmaterial}H. Kageyama, K. Yoshimura, R. Stern, N. V. Mushnikov, K. Onizuka, M. Kato, K. Kosuge, C. P. Slichter, T. Goto, and Y. Ueda, Phys. Rev. Lett. 82, 3168 (1999).

\bibitem{compound}Y. Tsujimoto, Y. Baba, N. Oba, H. Kageyama, T. Fukui, Y. Narumi, K. Kindo, T. Saito, M. Takano, Y. Ajiro, and K. Yoshimura, J. Phys. Soc. Jpn. 76, 063711 (2007).

\end{thebibliography}

\begin{thebibliography}{2}
\comment{
\bibitem{Arfken}G. Arfken, Mathematical Methods for Physicists $3^{\mathrm{rd}}$ edition (Associated Press, 1985).
}
\bibitem{AGDsuppmat}A. A. Abrikosov, L. P. Gorkov, and I. E. Dzyaloshinskii, Methods of Quantum Field Theory in Statistical Physics (Dover Publications, 1963).

\end{thebibliography}
\end{document}